\documentclass[prd,preprint,superscriptaddress,showpacs,byrevtex]{revtex4}
\usepackage{epsfig}

\newcommand{\be}{\begin{equation}}
\newcommand{\ee}{\end{equation}}
\newcommand{\bea}{\begin{eqnarray}}
\newcommand{\eea}{\end{eqnarray}}
\def\simgt{\rlap{\lower 3.5 pt \hbox{$\mathchar \sim$}} \raise 1pt
Ê \hbox {$>$}}

\begin{document}

\title{ String Theory at LHC Using Top Quarks From String Balls }

\author{Gouranga C. Nayak} \email{nayak@physics.arizona.edu}
\affiliation{ Department of Physics, University of Arizona, Tucson, AZ 85721 USA
}

\date{\today}

\begin{abstract}

According to string theory, string ball is a highly excited long string which decays
to standard model particles at the Hagedorn temperature with thermal spectrum. If
there are extra dimensions, the string scale can be $\sim$ TeV, and we should produce
string balls at CERN LHC. In this paper we study top quark production from string balls
at LHC and compare with the parton fusion results at NNLO using pQCD. We find
significant top quark production from string balls at LHC which is
comparable to standard model pQCD results. We also find that $\frac{d\sigma}{dp_T}$ of top
quarks from string balls does not decrease significantly with increase in $p_T$, whereas it
deceases sharply in case of standard model pQCD scenario. Hence, in the absence of black
hole production at LHC, an enhancement in top quark cross section and its abnormal
$p_T$ distribution can be a signature of TeV scale string physics at LHC. String theory is
also studied at LHC via string Regge excitations in the weak coupling limit in model independent
framework. Since massive quark production amplitude is not available in string Regge excitations scenario,
we compute massless quark production in string Regge excitations scenario and make a
clear comparison with that produced from string balls at LHC for a given luminosity.

\end{abstract}
\maketitle

\newpage

\section{Introduction}

It is now generally accepted that the scale of quantum gravity {\it could be} as low as one TeV \cite{folks,ppbf,pp,pp1,pp2,pp3,ag,ppch,ppk,ppu,park,hof,more,gram,cham,pp4,pp5,pp6}. In the
presence of extra dimensions, the string mass scale $M_s$ and the
Planck mass $M_P$ could be around $\sim$ TeV. In this situation we can
look forward to search for TeV scale string physics at
CERN LHC. One of the most exciting possibility is to search for TeV scale black hole
and string ball production at LHC. These `brane-world' black holes and string balls will
be our first window into the extra dimensions of space predicted by string theory,
and required by the several brane-world scenarios \cite{large}.
There may be many other ways of testing string theory at LHC
starting from from brane excitations to various string excitations. The
string balls of \cite{dimo} is just one such model, where the predictions
are done in a toy string theory model. In this paper we will focus on studying
string theory at LHC based on string balls.

There has been arguments that the black hole stops radiating near
Planck scale and forms a black hole remnant \cite{dasall}. These black hole
remnants can be a source of dark matter \cite{steinhardt, darknayak}. In the
absence of a theory of quantum gravity, we can study other scenarios
of black hole emissions near the Planck scale. Ultimately, experimental
data will determine which scenarios are valid near the Planck scale.
In this paper we will study string ball production at LHC in the context
of black hole evaporation in string theory. Recently, string theory has given
convincing microscopic calculation for black hole evaporation
\cite{susskind,allstring}.

String theory predicts that a black hole has formed at several times
the Planck scale and any thing smaller will dissolve into some thing
known as string ball \cite{dimo}. A string ball is a highly
excited long string which emits massless (and massive) particles at Hagedorn
temperature with thermal spectrum \cite{amati,canada}. For general relativistic
description of the back hole to be valid, the black hole mass $M_{BH}$ has to be
larger than the Planck mass $M_P$. In string theory the string ball mass $M_{SB}$
is larger than the string mass scale $M_s$. Typically
\bea
&& M_s < M_P < \frac{M_s}{g_s^2}  \nonumber \\
&& M_s << M_{SB} << \frac{M_s}{g_s^2} \nonumber \\
&& \frac{M_s}{g_s^2} << M_{BH}
\label{scale}
\eea
where $g_s$ is the string coupling which can be less than 1
for the string perturbation theory to be valid. Since string ball
is lighter than black hole, more string balls are expected to be
produced at CERN LHC than black holes.

The Hagedorn temperature of a string ball is given by
\bea
T_{SB}=\frac{M_s}{\sqrt{8} \pi}
\label{tsb}
\eea
where $M_s ~\sim $ TeV is the string scale. Since this temperature is very high at LHC
($\sim$ hundreds of GeV) we expect more massive particles ($M \sim 3 T_{SB}$) to be
produced at CERN LHC from string balls.

Top quark is the heaviest observed particle in the standard model. It was discovered
in $p\bar p$ collisions at $\sqrt s$ = 1.8 TeV at Tevatron.
Run II of Tevatron at $\sqrt s$ = 1.96 TeV has triggered more studies on top quark properties.
On the theoretical side, there have been progress on next-to-next leading
order (NNLO) pQCD calculations at LHC. Since LHC energy (pp collisions at
$\sqrt s$ = 14 TeV) is much larger than the Tevatron energy, LHC is expected
to be a top quark factory. This is because gluon-gluon fusion processes at
low $x$ are expected to dominate the top quark cross section (about 90 percent).

In this paper we study top quark production at CERN LHC from string balls and
make a comparison with the top quark production from the parton fusion processes
at NNLO using pQCD. We present the results for the total cross section and
$\frac{d\sigma}{dp_T}$ of top quarks. There can be significant top quark production
from black holes at LHC as well \cite{nayaktop}. This is because the black hole
temperature increases as its mass decreases whereas the string ball temperature
remains constant (see eq. (\ref{tsb})). On the other hand the string ball mass
is smaller than the black hole mass and more string balls are produced at LHC.
Hence top quark production at LHC is from two competitive effects:
1) string ball (black hole) production at LHC and
2) top quarks emission from a single string ball (black hole) at LHC.
We find that top quark production from string balls can be comparable
to that from black holes at LHC. Hence, in the absence of black hole production
at LHC, an enhancement in the top quark cross section may be a signature of TeV
scale string physics at LHC.

String theory is also studied at LHC via string Regge excitations in the weak coupling
limit in model independent framework \cite{dstring}. Since massive quark production amplitude is
not available in string Regge excitations scenario, we compute massless quark production
in string Regge excitations scenario and make a clear comparison with that produced from
string balls at LHC for a given luminosity.

The paper is organized as follows. In section II we discuss string ball
production and its decay in string theory and at LHC. In section III we discuss
top quark production from string balls at the CERN LHC. Section IV describes
top quark production in pQCD at NNLO. In section V we briefly describe low mass
string Regge excitations scenario in partonic collisions at LHC. We present our
results and discussions in section VI and conclude in section VII.

\section{ String Ball Production and Decay in String Theory and at LHC }

Fundamental scales used in string theory are as follows: $l_P$ is the Planck length
scale, $l_s$ is the quantum length scale of the string, $\alpha'=l_s^2$ is the
inverse of the classical string tension, $M_s=\frac{1}{l_s}$ is the string mass
scale and $g_s$ is the string coupling. For small string coupling
\bea
l_P \sim g_s l_s.
\label{length}
\eea
In $d=3+n$ space dimensions one obtains
\bea
l_P^{d-1} \sim g_s^2 l_s^{d-1}.
\label{dlength}
\eea

According to string theory as black hole shrinks it reaches the correspondence point
\cite{susskind,allstring}
\bea
M \le M_c \sim \frac{M_s}{g^2_s}
\eea
and makes a transition to a configuration dominated by a highly excited long string.
This highly excited long string (known as string ball) continues to lose mass by evaporation
at the Hagedorn temperature \cite{amati} and "puffs-up" to a larger "random-walk" size which
has observational consequences. Evaporation, still at the Hagedorn temperature,
then gradually brings the size of the string ball down towards $l_s$.

Production of a highly excited string from the collision of two
light string states at high $\sqrt{s}$ can be obtained from the Virasoro-Shapiro
four point amplitude by using string perturbation theory. One finds the amplitude
\bea
A(s,t)=\frac{2\pi g_s^2 \Gamma[-1-\alpha's/4]\Gamma[-1-\alpha't/4]\Gamma[-1-\alpha'u/4]}{\Gamma[2+\alpha's/4]
\Gamma[2+\alpha't/4]\Gamma[2+\alpha'u/4]}
\eea
with
\bea
s+t+u =   -16 /\alpha'.
\eea
The production cross section is
\bea
\sigma \sim \frac{\pi {\rm Res} A(\alpha's/4 = N, t=0)}{s} = g_s^2 \frac{\pi^2}{8} \alpha'^2 s.
\label{cs1}
\eea

The cross section in eq. (\ref{cs1}) saturates the
unitarity bounds at around $g_s^2 \alpha' s \sim 1$. This implies that the production cross section for
string balls grows with $s$ as in eq. (\ref{cs1}) only for
\bea
M_s << \sqrt{s} << M_s/g_s,
\eea
while for
\bea
\sqrt{s} >> M_s/g_s,~~~~~~~~~~~~~~~~~~~~~~~~~~\sigma_{SB} = \frac{1}{M_s^2}
\eea
which is constant.

Hence the string ball production cross section in a parton-parton collision is given by \cite{dimo}
\bea
&&~ \sigma_{SB} \sim \frac{g_s^2 M_{SB}^2}{M_s^4},~~~~~~~~~~~M_s<<M_{SB}<<M_s/g_s, \nonumber \\
&&~ \sigma_{SB} \sim \frac{1}{M_s^2},~~~~~~~~~~~M_s/g_s<<M_{SB}<<M_s/g^2_s.
\label{sbc}
\eea

Highly excited long strings emit massless (and massive) particles
at Hagedorn temperature \cite{amati}. Hence the
conventional description of evaporation in terms of black body radiation
can be applied. The emission can take place either in the bulk (in to
the closed string) or in the brane (in to open strings). The wavelength
\bea
\lambda =\frac{2\pi}{T_{SB}}
\eea
corresponding to Hagedorn temperature is larger than the size of the string ball. So the compact string ball
is, to first approximation, a point radiator and, consequently, emits mostly s-waves. This indicates that
it decays equally to a particle on the brane and in the bulk. This is because s-wave emission is sensitive
only to the radial coordinate and does not make use of the extra angular modes available on the bulk.
Since there are many more species of particles ($\sim$ 60) on our brane than our bulk, the string ball
decays visibly to standard model particles \cite{bv,dimo}.

However, when string ball puffs-up to a larger random walk size, its spatial extent can approach
or exceed the wavelength of the emitted quanta, which implies that it can use more of the angular
modes that the additional dimensions provide. The average radius of the string ball is
\bea
R_{SB} \sim l_s\sqrt{M_{SB}l_s}
\label{rsb}
\eea
This, however, is a temporary effect: as the
string ball decays, its size shrinks towards $l_s$, and once again, it becomes a small
radiator emitting mostly at brane.

\section{ Top Quark Production From String Balls at LHC }

If string balls are formed at the LHC then they will quickly evaporate
by emitting massless (and massive) particles at Hagedorn temperature
with thermal spectrum \cite{amati,canada}.
The emission rate for top quark with mass $M_t$, momentum $\vec{p}$ and energy
$E =\sqrt{{\vec p}^2+M_t^2}$ from a string ball of temperature $T_{SB}$
is given by
\be
\frac{dN}{d^3p dt }=
\frac{c_s \sigma_s}{(2\pi)^3}\frac{1}{(e^{\frac{E}{T_{SB}}} + 1)}\,,
\label{thermal}
\ee
where $\sigma_s$ is the area factor \cite{bv} and $c_s=6$ is the multiplicity factor
for top quark.

Note that we do not assume B and L or B-L conservation in this paper. Depending on the
assumptions made the probability of the top-quark emission change. In one case the
top quark has to be emitted together with an antiquark and in other case it needs
to be accompanied by either antiquark or a lepton.  In our calculation we do not
assume any quantum number conservation. The case of no conservation at all would
violate many known bounds on baryon and lepton number conservation.
Since we will compare our results with the pQCD NNLO computation at LHC
\cite{vogt,vogt1} which assumes no quantum number conservation we will assume no quantum
number conservation in this paper.

This result in Eq. ({\ref{thermal}}) is for top quark production from a single
string ball of temperature $T_{SB}$. To obtain total top quark cross section
at LHC we need to multiply the number of top quarks produced from a
single string ball with the total string ball production cross section in
pp collisions at LHC.

The string ball production cross section in pp collisions at $\sqrt{s}$= 14 TeV at
LHC is given by \cite{pp,cham},
\bea
\sigma_{SB}^{pp \rightarrow SB +X}(M_{SB})
= {\sum}_{ab}~
\int_{\tau}^1 dx_a \int_{\tau/x_a}^1 dx_b f_{a/p}(x_a, Q^2)
\nonumber \\
\times f_{b/p}(x_b, Q^2)
\hat{\sigma}^{ab \rightarrow SB }(\hat s) ~\delta(x_a x_b -M_{SB}^2/s).
\label{bkt}
\eea
In this expression $\hat{\sigma}^{ab \rightarrow SB }(\hat s)$
is the string ball cross section in partonic collisions which is given by eq. (\ref{sbc}),
$x_a (x_b)$ is the longitudinal momentum fraction of the parton inside
the hadron A(B) and $\tau=M_{SB}^2/s$. Energy-momentum conservation implies
$\hat s =x_ax_b s=M_{SB}^2$. We use $Q = M_{SB}$ as the factorization scale at which the
parton distribution functions are measured. ${\sum}_{ab}$ represents
the sum over all partonic contributions where $a,b=q, {\bar q}, g$. We use CTEQ6 \cite{cteq}
PDF to compute the string ball cross section at LHC.

The total top quark production cross section in the process
$pp \rightarrow SB+X$  at LHC is then given by
\be
\sigma = N \times \sigma_{SB}
\label{sbtp}
\ee
where $\sigma_{\rm SB}$ is given by eq. (\ref{bkt}). To obtain $p_T$
distribution we use $d^3p~=~2\pi ~dp_T ~p_T^2 ~dy ~{\rm cosh}y $ in
eq. (\ref{thermal}) where $y$ is the rapidity.

\section{ Top Quark Production via pQCD Processes at the LHC}

The top quarks at LHC are mainly produced in $t\bar t$ pairs.
At the LHC proton-proton collider, the QCD production process
involves quark-antiquark and gluon-gluon fusion mechanism.
The gluon-gluon fusion processes give the dominant cross section
(about 90 percent). This subprocess at high energy is the main reason
for larger rate of the cross section compared to Tevatron at Fermilab.
The single top quark production occurs via electroweak process. The single
top quark production cross section ($\sim $ 300 pb ) is smaller
compared to $t\bar t$ total cross section ($\sim 970$ pb) at LHC at
$\sqrt{s}$ =14 TeV pp collisions. Hence we will not consider the single top quark production cross section
\cite{sing} in this paper. We will consider $t \bar t$ pair production using
parton fusion processes at LHC and will
compare them with the top quark production cross section from string balls.

At the next-to-next-to-leading order (NNLO) one needs to compute the
following partonic subprocesses. On the leading-order (LO) level we have
\bea
q + {\bar q} \rightarrow t {\bar t},~~~~~~~~~~~g + g \rightarrow t {\bar t}.
\label{LO}
\eea
In NLO we have in addition to the one-loop virtual corrections to the above
reaction the following two-to-three body processes
\bea
q + {\bar q} \rightarrow t {\bar t}+g,~~~~~~~~~~~g+q ({\bar q}) \rightarrow t {\bar t}+q({\bar q}),
~~~~~~~~~~~g + g \rightarrow t {\bar t}+g.
\label{NLO}
\eea
At NNLO level we receive the two-loop virtual corrections to the LO processes in eq. (\ref{LO})
and one-loop virtual corrections to NLO reactions in eq. (\ref{NLO}). To these contribution
one has to add the results obtained from the following two-to-four body reactions
\bea
&& g + g \rightarrow t {\bar t}+g+g, ~~~~~~~~~~~~~~~g+g \rightarrow t{\bar t}+q+ {\bar q}, \nonumber \\
&& g+q ({\bar q}) \rightarrow t {\bar t}+q({\bar q})+g, \nonumber \\
&& q + {\bar q} \rightarrow t {\bar t}+g+g, ~~~~~~~~~~~~~~~q+{\bar q} \rightarrow t{\bar t}+q+ {\bar q}, \nonumber \\
&& q +  q \rightarrow t {\bar t}+q+q, ~~~~~~~~~~~~~~~{\bar q}+{\bar q} \rightarrow t{\bar t}+{\bar q}+ {\bar q}, \nonumber \\
&& q_1 +  q_2 \rightarrow t {\bar t}+q_1+q_2, ~~~~~~~~~~~~~~~q_1+ {\bar q}_2 \rightarrow t{\bar t}+q_1+  {\bar q}_2. \label{NNLO}
\eea
After the phase space integrals has been done the partonic cross section ${\hat \sigma}$ is
rendered finite by coupling constant renormalization, operator renormalization and the removal of collinear divergences. The renormalization scale $\mu_R$ is set equal to the mass
factorization scale $\mu_F$. The cross section
for top quark production in proton-proton collisions at the LHC
is given by
\begin{eqnarray}
\label{eqn4.1}
d\sigma=\sum_{a,b=q,\bar q,g} \int dx_1 \int dx_2\,f_a(x_1,\mu_F^2)
f_b (x_2,\mu_F^2 ) ~d{\hat \sigma}_{ab}
\end{eqnarray}
where $d{\hat \sigma}_{ab}$ is the partonic level differential cross section for top quark production.
For the details, see \cite{toplhc1,vogt,vogt1}. Reviews of present status of top quark physics
at LHC can be found in \cite{toplhc2}.

\section{ Parton Production in $ 2 \rightarrow 2$ Processes via String Regge Excitations }

If the string mass scale $M_s \sim $ TeV, we can also expect to discover string Regge excitations with masses
of order $M_s$ in $2 \rightarrow 2$ partonic processes at LHC in the weak coupling limit in a model independent
framework \cite{dstring}. In this case a whole tower of infinite string excitations
will open up and the new particles follow the well known Regge trajectories of vibrating
string
\bea
j= j_0 + \alpha' M^2
\label{j}
\eea
with spin $j$. These stringy states will lead to new contributions to standard model
scattering processes. This is based on the extensions of standard model where open strings
ends on D-branes, with gauge bosons due to strings attached to stacks of D-branes and
chiral matter due to strings stretching between intersecting D-branes \cite{dbrane}.

Dijet production in the string resonance scenario in partonic collisions
at LHC is studied in \cite{dstring}. The $ 2 \rightarrow 2$ partonic scattering
amplitudes are computed at the leading order in string perturbation theory \cite{2string}.
In this calculation ${\hat s}+{\hat t} +{\hat u}=0$ is used which is the case for massless partons in the initial and
final states in the $2 \rightarrow 2$ partonic scattering processes. Here ${\hat s}$, ${\hat t}$ and ${\hat u}$
are the Mandelstam variables at partonic level. For top quark production one needs to extend
this string Regge formalism to the case ${\hat s}+ {\hat t} + {\hat u} \neq 0$ to calculate the amplitude.
Since massive
quark production amplitude is not available in string Regge excitations scenario, we will compute
massless quark production in string Regge excitations scenario and will make a clear comparison with
that produced from string balls at LHC for a given luminosity.
It can be mentioned that a similar situation exists in AdS/CFT scenario as well, where gluon scattering
amplitude \cite{alday} and massless quark scattering amplitudes are studied \cite{mit}. The
partonic scattering amplitude for the massive quark production in the final state in the
AdS/CFT scenario is expected to be complicated and has not been studied so far.

Since the gluon fusion process is dominant at LHC we will consider the process $gg \rightarrow q \bar q$ via
string Regge excitation in this paper. The matrix element square for this process is given by \cite{dstring}
\bea
&&~|M(gg \rightarrow q \bar q)|^2 = \frac{7}{24} \frac{16 \pi^2 \alpha_s^2}{M_s^4} N_f \times \nonumber \\
&&~[W_{g^*}^{gg \rightarrow q \bar q} \frac{{\hat u}{\hat t}({\hat u}^2+{\hat t}^2)}{({\hat s}-M_s^2)^2+(\Gamma_{g^*}^{J=2} M_s)^2}
+W_{C^*}^{gg \rightarrow q \bar q} \frac{{\hat u}{\hat t}({\hat u}^2+t^2)}{({\hat s}-M_s^2)^2+(\Gamma_{C^*}^{J=2} M_s)^2}]
\label{m2}
\eea
where $\alpha_s$ is the QCD coupling constant and
\bea
&&~W_{g^*}^{gg \rightarrow q \bar q}=0.24,~~~~~~~~~~~~~~~W_{C^*}^{gg \rightarrow q \bar q}=0.76, \nonumber \\
&&~\Gamma_{g^*}^{J=2}=45 (M_s/{\rm TeV}) {\rm GeV},~~~~~~~~~~\Gamma_{C^*}^{J=2}=75 (M_s/{\rm TeV}) {\rm GeV}.
\label{rest}
\eea
The differential cross section for jet production in pp collisions at LHC is given by
\bea
E\frac{d\sigma}{d^3p} =  \int dx_1 \int dx_2 f(x_1, Q^2) f(x_2,Q^2) \frac{\hat s}{\pi}
\frac{d{\hat \sigma} }{d{\hat t}} \delta({\hat s} +{\hat t} + {\hat u})
\label{eds}
\eea
where $\frac{d{\hat \sigma} }{d{\hat t}}$ is the partonic level differential cross section.
This gives for the quark jet production
\bea
\frac{d\sigma}{dp_T} = \frac{p_T}{{8 \pi s}}~\int dy \int dy_2 ~\frac{1}{{\hat s}}~f_g(x_1, Q^2)~ f_g(x_2,Q^2)~\times~
|M(gg \rightarrow q \bar q)|^2
\label{dsdpt}
\eea
where
\bea
x_1=\frac{p_T}{\sqrt{s}} [e^{y}+e^{y_2}],~~~~~~~~~~~~~~x_2=\frac{p_T}{\sqrt{s}} [e^{-y}+e^{-y_2}].
\label{x12}
\eea
$|M(gg \rightarrow q \bar q)|^2$ is given by eq. (\ref{m2}) for the process $gg \rightarrow q \bar q$ in the string Regge
excitation scenario \cite{dstring}. We will compare eq. (\ref{dsdpt}) with $\frac{d\sigma}{dp_T}$ from eq. (\ref{sbtp})
for massless quark production from string balls at LHC. We have used CTEQ6 PDF \cite{cteq} in our calculation.

\section{Results and Discussions}

In this section we will compute the top quark production
cross section from string balls at $\sqrt{s}$ = 14 TeV in pp collisions
and will compare them with the top quark production via parton fusion
processes at NNLO. The top quark production from string ball
is described in section III. For the string ball production
we choose the factorization and normalization scale to be
the mass of the string ball. As the temperature of the string ball
is very large there is not much difference in the top quark
production cross section from string balls if the top quark mass $M_t$
is increased from 165 to 180 GeV. String ball mass $M_{SB}$ should be
larger than the string scale $M_S$. We take
\bea
M_{SB} \ge 3 M_s,~~~~~~~~~~~~~~~~~~g_s=0.3.
\eea
in our calculation \cite{canada}.

\begin{figure}[htb]
\vspace{2pt}
\centering{{\epsfig{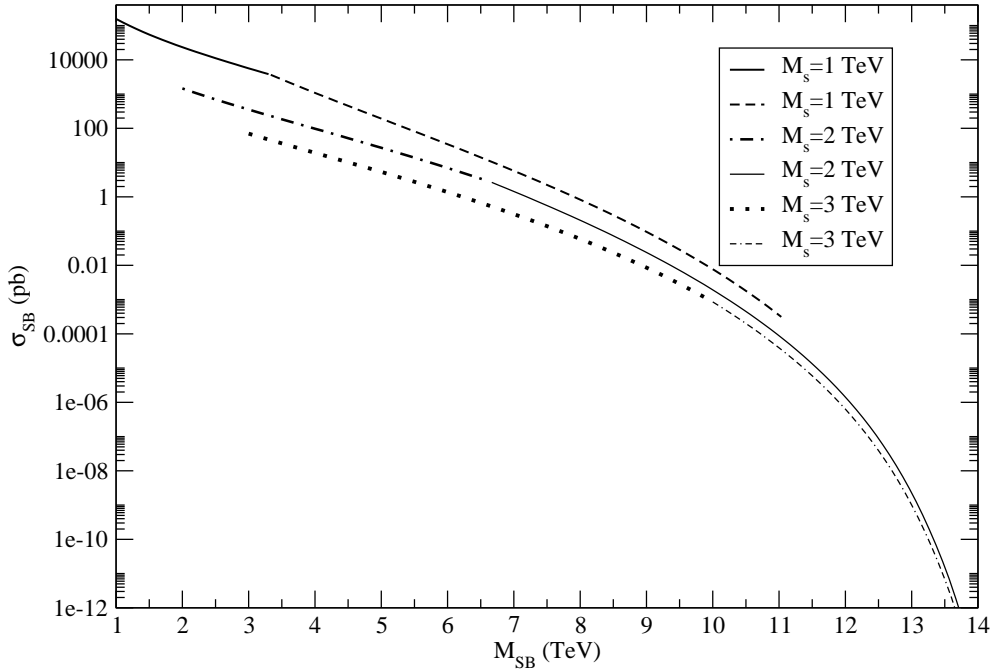}}}
\caption{ Total cross sections for string ball production at the LHC.}
\label{fig1}
\end{figure}

In Fig. 1 we present the string ball production cross section
at the LHC.  The $y$-axis
is the string ball production cross section $\sigma_{SB}$ in
pb and the $x$-axis is the string ball mass $M_{SB}$ in TeV.
The upper curve is for string scale $M_s$ = 1 TeV. The solid
line in the upper curve is for the string ball mass in the range
$   M_s<<M_{SB}<<M_s/g_s$ and the dashed line is for
$M_s/g_s<<M_{SB}<<M_s/g^2_s$, see eq. (\ref{sbc}).
The middle curve is for string scale $M_s$ = 2 TeV. The dot-dashed
line in the middle curve is for the string ball mass in the range
$   M_s<<M_{SB}<<M_s/g_s$ and the solid line in the middle curve is for
$M_s/g_s<<M_{SB}<<M_s/g^2_s$.
The lower curve is for string scale $M_s$ = 3 TeV. The dotted
line in the lower curve is for the string ball mass in the range
$   M_s<<M_{SB}<<M_s/g_s$ and the dot-dashed-dashed line in the lower
curve is for $M_s/g_s<<M_{SB}<<M_s/g^2_s$.
As can be seen from the figure the cross sections decrease rapidly when
both the string mass scale $M_s$ and string ball mass $M_{SB}$ increases.
These string ball production cross sections will be
multiplied with the number of top quarks produced from a
single string ball to obtain the top quark production cross section
from string balls at the CERN LHC.

\begin{figure}[htb]
\vspace{2pt}
\centering{{\epsfig{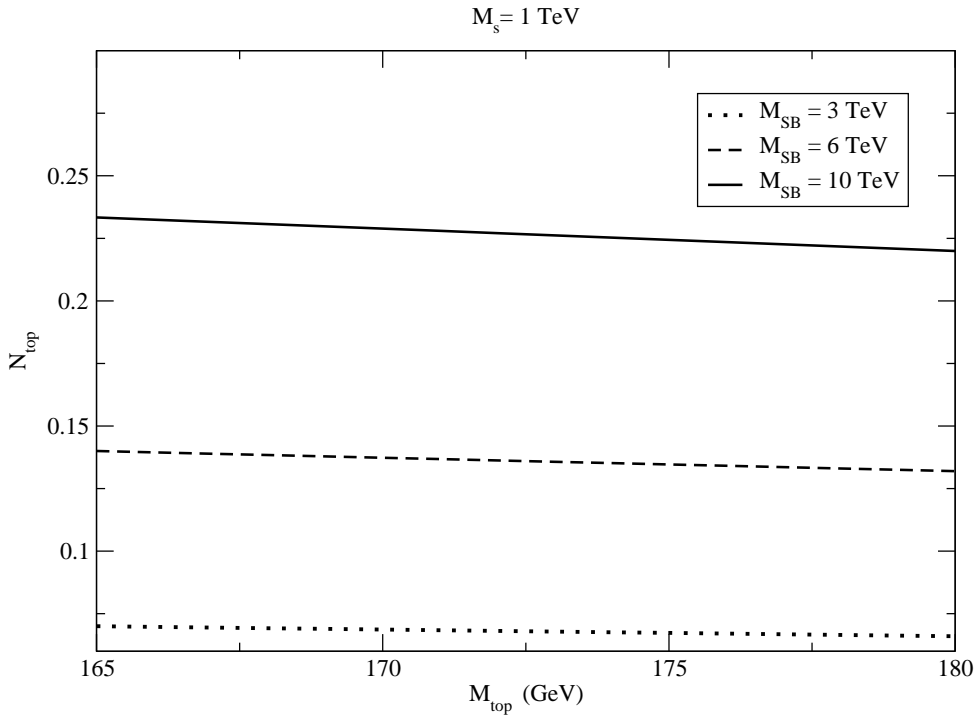}}}
\caption{ Average Number of top quark production from a single string ball
at LHC. }
\label{fig2}
\end{figure}

In Fig. 2 we present results for the average number of top quarks produced
from a single string ball as a function of top quark mass.
The $y$-axis is the average number of top quark production
from a single string ball and the $x$-axis is the mass of the top quark in GeV.
The solid, dashed and dotted lines are for string ball masses equal to 10,
6 and 3 TeV respectively. Unlike black hole case, the average number of top quarks
produced from a string ball is larger for larger mass string balls. This is
because the Hagedorn temperature, eq. (\ref{tsb}), is independent of the string ball
mass whereas the black hole temperature decreases as black hole mass increases. The increase
in number of top quarks is due to the increase in area of the string ball, see eqs.
(\ref{rsb}) and (\ref{thermal}). This is the case for emission from a single string ball.
The string ball production cross section itself decreases at LHC as the
mass of the string ball increases, see Fig.1. Hence the total cross section
of top quark production from string balls at LHC is a competitive effect
from the above two factors (see eq. (\ref{sbtp})).

\begin{figure}[htb]
\vspace{2pt}
\centering{{\epsfig{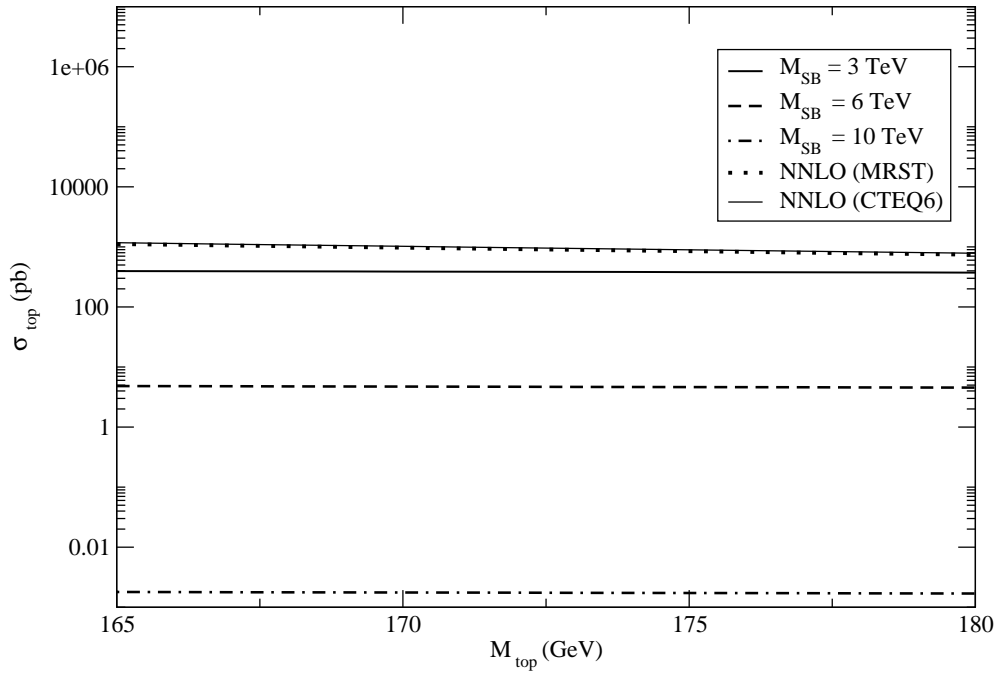}}}
\caption{ Total cross section for top quark production at LHC
from string balls and from direct pQCD processes at NNLO.
}
\label{fig3}
\end{figure}

In Fig.3 we present total top quark production cross section
from string balls and compare them with the pQCD predictions at NNLO.
The former is given for three different choices of the string ball mass, namely
$M_{SB}$= 3, 6 and 10 TeV respectively. We plot for comparison the NNLO
top quark cross section with $\mu_F=\mu_R=M_t$ \cite{vogt}.
The dotted line for NNLO curve is for MRST 2006 PDF and the thin-solid line for NNLO curve
is for CTEQ6.6 PDF. The thick-solid line is for string ball mass equal to 3 TeV, the dashed
line is for string ball mass equal to 6 TeV and the dot-dashed line is for
string ball mass equal to 10 TeV. For larger value of string mass scale $M_s$
the cross section becomes even smaller and hence we do not plot them. It is clear that
the total top quark cross section from string balls is comparable to pQCD cross section
for small value of string mass scale ($M_s\sim $ 1 TeV) and string ball mass
($M_{SB}\sim $ 3 TeV) and is not sensitive to the increase in top quark mass $M_t$.

\begin{figure}[htb]
\vspace{2pt}
\centering{{\epsfig{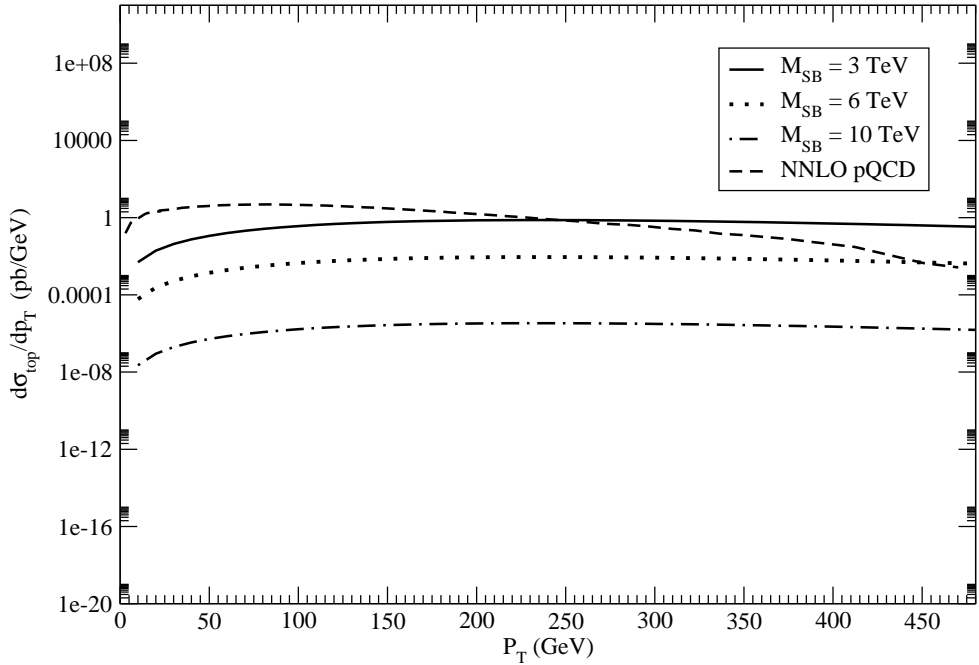}}}
\caption{ Transverse momentum distribution of top quark production at LHC
from string balls and from direct pQCD processes at NNLO.}
\label{fig4}
\end{figure}

In Fig.4 we present $\frac{d\sigma}{dp_T}$ of top quark production
from string balls at LHC and compare them with the pQCD predictions at NNLO.
The top quark mass is chosen to be 175 GeV. The $\frac{d\sigma}{dp_T}$ of top
quark from string balls is given for three different choices of the string ball mass,
namely $M_{SB}$= 3, 6 and 10 TeV respectively with string mass scale $M_s$ = 1 TeV
in each case. We plot for comparison the NNLO results
for $\frac{d\sigma}{dp_T}$ of top quark using pQCD \cite{vogt1}. The
dashed line is the NNLO pQCD result. The solid line is for string ball mass equal to 3
TeV, the dotted line is for string ball mass equal to 6 TeV and the
dot-dashed line is for string ball mass equal to 10 TeV. For larger
value of string scale $M_s$ the cross section becomes even smaller and hence we do not plot them.
It is clear that the $\frac{d\sigma}{dp_T}$ of the top quark
via string ball production is larger than the standard model pQCD predictions
for larger values of $p_T$ ($p_T>$ 250 GeV) of top quark and for smaller value
of string ball mass ($M_{SB} \sim$ 3 TeV). For larger values of $p_T$, the
$\frac{d\sigma}{dp_T}$ of top quark from string balls does not sharply decrease
with increasing $p_T$, whereas in case of NNLO pQCD processes it decreases sharply.
This fact can be used to distinguish between top quark production from string
balls and from parton fusion processes at NNLO in pQCD at LHC.

\begin{figure}[htb]
\vspace{2pt}
\centering{{\epsfig{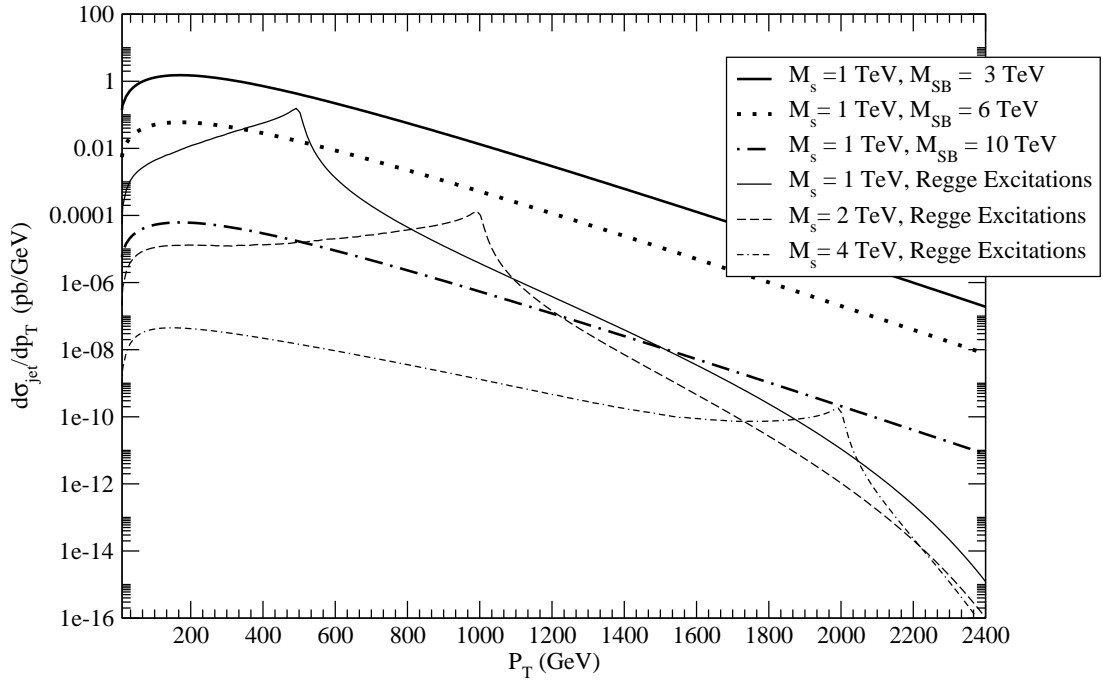}}}
\caption{ Transverse momentum distribution of quark jet production from string
balls and from string Regge excitations at LHC.}
\label{fig5}
\end{figure}

Since massive
quark production amplitude is not available in string Regge excitations scenario,
in Fig. 5 we present $\frac{d\sigma}{dp_T}$ of massless quark production from string balls
and compare with the process $gg \rightarrow q {\bar q}$ at LHC in the string Regge excitation
scenario \cite{dstring} by using eq. (\ref{dsdpt}). We have considered single quark flavor which corresponds
to the massless limit of the top quark. See section V for details. The thick solid line is for string mass
scale $M_s$ = 1 TeV and string ball mass $M_{SB}$= 3 TeV. The dotted line is for $M_s$ = 1 TeV and $M_{SB}$= 6 TeV.
The dot-dashed line is for $M_s$ = 1 TeV and $M_{SB}$= 10 TeV. For comparison we present $\frac{d\sigma}{dp_T}$
of massless quark production in the string Regge excitation scenario. The thin solid, dashed and dot-dashed-dashed
lines are in the string Regge excitation scenario for $M_s$ = 1, 2 and 4 TeV respectively. It can be seen
that the resonances are observed in case of string excitation scenario which is absent in the string ball
scenario.

In Fig. 6 we present number of quark jets per GeV from string balls
at LHC and compare with the process $gg \rightarrow q {\bar q}$ at LHC in the string Regge excitation
scenario \cite{dstring} with the luminosity of 10 ${\rm pb}^{-1}$.
The thick solid line is for string mass
scale $M_s$ = 1 TeV and string ball mass $M_{SB}$= 3 TeV. The dotted line is for $M_s$ = 1 TeV and $M_{SB}$= 6 TeV.
The dot-dashed line is for $M_s$ = 1 TeV and $M_{SB}$= 10 TeV. For comparison we present number of quark jets per GeV
in the string Regge excitation scenario. The dot-dashed-dashed and dot-dot-dashed lines
are in the string Regge excitation scenario for $M_s$ = 1 and 2 TeV respectively.
We also present the results of the QCD jets at the CMS detector at LHC. The thin solid line
is the result for CMS QCD jets taken from \cite{qcdjet} with the same luminosity of 10 ${\rm pb}^{-1}$. It can be seen
that the resonances are observed in case of string excitation scenario which is absent in the string ball
scenario and in QCD jets scenario.

\begin{figure}[htb]
\vspace{2pt}
\centering{{\epsfig{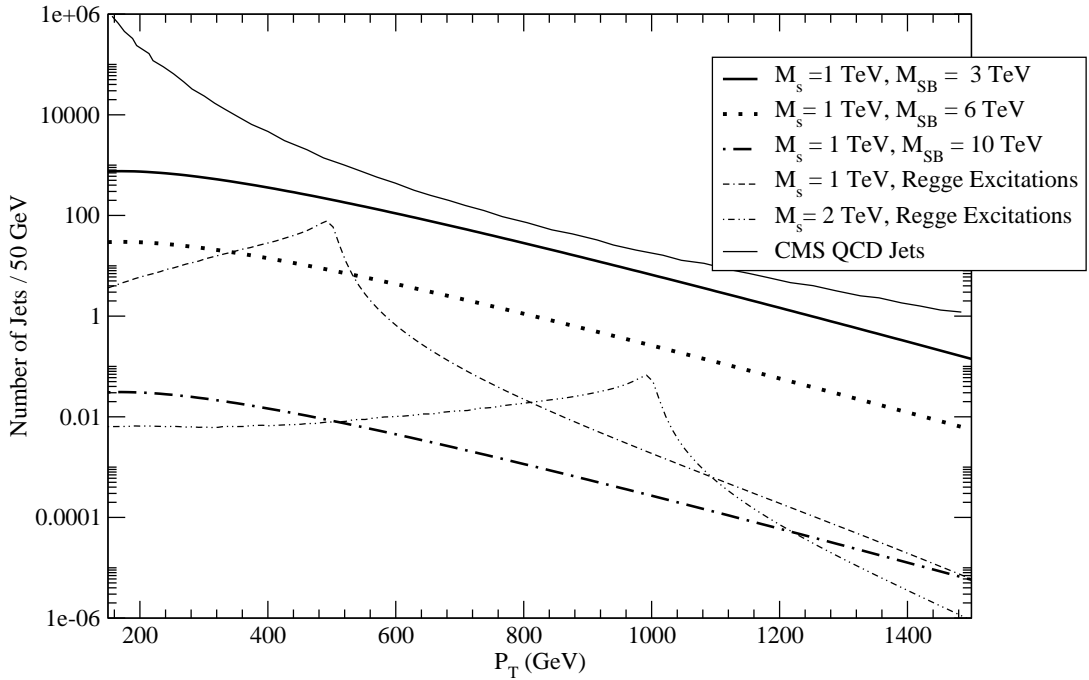}}}
\caption{ Number of quark jets per GeV (in 50 GeV $p_T$ bin) as a function of $p_T$ from string balls and from
string Regge excitations at LHC. The luminosity used is 10 ${\rm pb}^{-1}$. CMS QCD jets are from \cite{qcdjet}. }
\label{fig6}
\end{figure}

\section{Conclusions}

In summary, we have computed top quark production cross section and its
$p_T$ distribution from string balls in proton-proton collisions at the LHC at $\sqrt{s}$ = 14 TeV
in the context of string theory and TeV scale gravity. We have compared the result with the pQCD cross
sections at NNLO. As the temperature of the string ball is large there is a huge amount of top quark
production from string balls at the LHC if the string scale is $\sim$ 1 TeV and the string ball
mass is $\sim$ 3 TeV. We have found that, unlike standard model predictions,
the top quark production cross section from string ball is not sensitive to the increase in top quark mass.
We have also found that $\frac{d\sigma}{dp_T}$ of top quark from
string balls does not sharply decrease with increasing $p_T$,
whereas in standard model processes it decreases sharply. Hence, in the
absence of black hole production at LHC, an enhancement in top quark cross
section and its abnormal $p_T$ distribution can be a signature of TeV scale
string physics at LHC.

String theory is also studied at LHC via string Regge excitations in the weak coupling
limit in model independent framework \cite{dstring}. Since massive
quark production amplitude is not available in string Regge excitations scenario, we have computed
massless quark production in string Regge excitations scenario and have made a clear comparison with
that produced from string balls at LHC for a given luminosity.

String balls might also be produced in PbPb collisions with a larger rate
\cite{cham}. In this case there can be additional effects of quark-gluon
plasma \cite{qgp} on the string ball radiation \cite{amati}.

\acknowledgments

This work was supported in part by Department of Energy under
contracts DE-FG02-91ER40664, DE-FG02-04ER41319 and DE-FG02-04ER41298.

\end{document}